\documentclass[useAMS,usenatbib]{mn2e}
\usepackage{amsmath,amssymb,amsfonts,bm}
\usepackage{natbib}
\usepackage{graphicx}
\usepackage{hyperref}
\usepackage{calc}
\usepackage{color}
\usepackage{wrapfig}
\usepackage{url}
\usepackage{paralist}
\usepackage{array}
\usepackage{tabularx}
\usepackage{txfonts}

\title[Mass discrepancy in galaxy clusters]{Mass discrepancy in galaxy clusters as a result of the offset between dark matter and baryon distributions}

\author[HuanYuan Shan et al.]{HuanYuan Shan$^{1,2}$\thanks{E-mail:
shanhuany@gmail.com, qinbo@bao.ac.cn}, Bo Qin$^{2, \star}$, and HongSheng Zhao$^{2,3}$\\
$^{1}$Department of Astronomy, School of Physics, Peking University,
Beijing, 100871, China\\
$^{2}$National Astronomical Observatories, Chinese Academy of Sciences, Beijing 100012, China\\
$^{3}$SUPA, University of St Andrews, KY16 9SS, UK}

\begin{document}

\date{Accepted \dots . Received \dots; in original form \dots}


\maketitle

\label{firstpage}

\begin{abstract}
   Recent studies of lensing clusters reveal that
it might be fairly common for a galaxy cluster that the X-ray center
has an obvious offset from its gravitational center which is
measured by strong lensing. We argue that if these offsets exist,
then X-rays and lensing are indeed measuring different regions of a
cluster, and may thus naturally result in a discrepancy in the
measured gravitational masses by the two different methods. Here we
investigate theoretically the dynamical effects of such
lensing-X-ray offsets, and compare with observational data. We find
that for typical values, the offset alone can give rise to a factor
of two difference between the lensing and X-ray determined masses
for the core regions of a cluster, suggesting that such ``offset
effect'' may play an important role and should not be ignored in our
dynamical measurements of clusters.
\end{abstract}


\section{Introduction}

Galaxy clusters, the largest gravitationally-bound structures in the
universe, are ideal cosmological tools. Accurate measurements of
their masses provide a crucial observational constraint on
cosmological models. Several dynamical methods have been available
to estimate cluster masses, such as (1) optical measurements of the
velocity dispersions of cluster galaxies, (2) measurements of the
X-ray emitting gas, and (3) gravitational lensing. Good agreements
between these methods have been found on scales larger than cluster
cores.

However, joint measurements of lensing and X-rays often identify
large discrepancies in the gravitational masses within the central
regions of clusters by the two methods, and the lensing mass has
always been found to be $2-4$ times higher than the X-ray determined
mass. This is the so-called ``Mass Discrepancy Problem'' (Allen
1998; Wu 2000). Many plausible explanations have been suggested,
e.g., the triaxiality of galaxy clusters (Morandi et al. 2010),
the oversimplification of the strong lensing model for the
central mass distributions of clusters (Bartelmann \& Steinmetz
1996), the inappropriate application of the hydrostatic equilibrium
hypothesis for the central regions of clusters (Wu 1994; Wu \& Fang
1997), or the magnetic fields in clusters (Loeb \& Mao 1994).

Recently Richard et al. (2010) present a sample of $20$ strong lensing
clusters taken from the Local Cluster Substructure Survey (LoCuSS),
among which $18$ clusters have X-ray data from Chandra observations
(Sanderson et al. 2009). They show that the X-ray/lensing mass
discrepancy is $1.3$ at $3\sigma$ significance --- clusters with
larger substructure fractions show greater mass discrepancies, and
thus greater departures from hydrostatic equilibrium.

On the other hand, lensing observations of the bullet cluster
1E0657-56 (Clowe et al. 2006), combined with earlier X-ray
measurements (Markevitch et al. 2006), clearly indicate that the
gravitational center of the cluster has an obvious offset from its
baryonic center. Furthermore, recent studies (Shan et al. 2010) of
lensing galaxy clusters reveal that offset between the lensing
center and X-ray center appears to be quite common, especially for
unrelaxed clusters. Among the recent sample of 38 clusters of Shan
et al. (2010), $45\%$ have been found to have offsets greater than
$10''$, and $5$ clusters even have offsets greater than $40''$.
Motivated by such observations, we propose to investigate galaxy
cluster models where the center of the dark matter (DM) halo does
not coincide with the center of the X-ray gas (See Figure~1).

\begin{figure}
\begin{center}
\protect\label{fig:offset}
\includegraphics[width=6.cm]{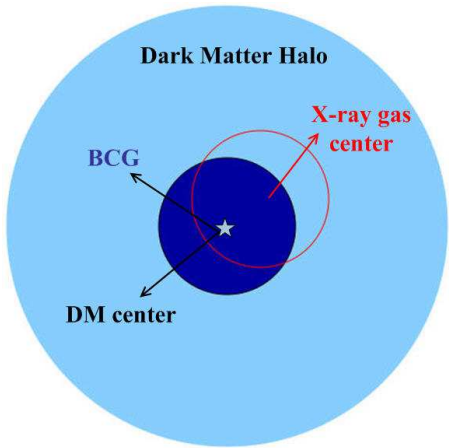}
\caption{Offset between the dark matter center and the X-ray center
in a galaxy cluster.}
\end{center}
\end{figure}

If the X-ray center of a cluster has an offset from its lensing
(gravitational) center, then the X-rays and lensing are indeed
measuring different regions of the cluster. Given the same radius,
the lensing is measuring the DM halo centered at the gravitational
center (shown by the dark blue sphere in Figure~1 ), while the
X-rays are measuring the sphere of the halo that is offset from the
{\it true} gravitational center (shown by the red circle in
Figure~1). In this case, there will always be a {\it natural}
discrepancy between the lensing and X-ray measured masses --- or
specifically, the X-ray mass will always be lower than the lensing
mass, just as the long-standing ``mass discrepancy problem'' has
indicated.

In this paper, we investigate the lensing-X-ray mass
discrepancy caused by the offsets between DM and X-ray gas. To check
our predictions, we compile a sample of $27$ clusters with good
lensing and X-ray measurements. We conclude that such ``offset''
effect should not be ignored in our dynamical measurements of galaxy
clusters. A flat $\Lambda$CDM cosmology is assumed throughout this
paper, where $\Omega_m$=0.3, $\Omega_{\Lambda}$=0.7, and
$\rm H_0=70 \, km \, s^{-1}Mpc^{-1}$.

\section{Mass discrepancy as a result of the dark matter-baryon offset}

We model our galaxy cluster with a fiducial model as the following:
(1) the DM halo is modeled by the Navarro-Frenk-White (NFW) profile
(Navarro et al. 1997) with concentration $c=4.32$ and scaled radius
$r_s=516 \, {\rm kpc}$, (2) the gas distribution is modeled by a
$\beta$ model with $\beta = 0.65$, the cluster core radius $r_c =
200 \, {\rm kpc}$, and the gas fraction $f_{\rm gas}=12\%$, (3) the
mass density of the BCG is described by a Singular Isothermal Sphere
(SIS) with a velocity dispersion of $300\, \rm km/s$.

The projected mass within a sphere of radius $R_{x}$ is
\begin{eqnarray*}
  m(R_{x},d) & = & \int_0^{2\pi} \int_0^{R_x}
                    \left[\Sigma_{\rm NFW}(R')+\Sigma_{\rm gas}(R) \right. \\
    & &
                    \left. + \Sigma_{\rm BCG}(R')\right]
                    R' \, dR' \, d{\theta},
\end{eqnarray*}
where $R'=\sqrt{d^2+R^2+2dR\cos \theta}$ is the 2-D radius from the
halo center, $R$ is the 2-D radius from the X-ray gas center, $d$ is
the 2-D offset between the halo center and X-ray center, and
$\Sigma_{\rm NFW}$, $\Sigma_{\rm gas}$, and $\Sigma_{\rm BCG}$ are
the projected mass densities of the DM halo, the gas and the BCG,
respectively. For a given radius $R_x$, the gravitational mass
measured by lensing $m_{\rm lens}$ can be given by $m(R_x,0)$ (as
shown by the dark blue sphere in Figure~1), while the projected mass
measured by X-rays $m_{\rm xray}$ is described by $m(R_x,d)$ (the
mass within the red circle in Figure~1). We now calculate the mass
ratio $m(R_x,d)/m(R_x,0)$, or equivalently, $m_{\rm lens}/m_{\rm
xray}$.

Figure~2 shows the mass ratio as a function of the 2-D offset $d$,
for a typical rich cluster. The solid curves are the mass ratio
with the fiducial model, the dashed and dotted curves are the mass ratio
with the NFW concentration $c=4.04$ and $5.13$ (top left), the
cluster core radius $r_c = 150 \, {\rm kpc}$ and $400 \, {\rm kpc}$
(top right), the $\beta$ index $\beta=0.6$ and $0.9$ (bottom left),
the gas fraction $f_{\rm gas}=0.1$ and $0.2$, respectively. For these
cases, the three curves from top to bottom are for the three measuring
radii $R_x=50\,{\rm kpc}, 100\,{\rm kpc}, 200\,{\rm kpc}$, respectively.
From Figure~2 we have the following conclusions:

(1) The lensing measured mass $m_{\rm lens}$ is always higher than
the X-ray measured mass $m_{\rm xray}$. For typical values of offset
$d=100\,{\rm kpc}$ and $R_x=100\,{\rm kpc}$, $m_{\rm lens}/m_{\rm
xray} \sim 2$, comparable to the ratio found in early studies (Allen
1998; Wu 2000; Richard et al. 2010).

(2) The ``Offset Effect'' we are reporting here should contribute
significantly to the long-standing ``Mass Discrepancy Problem''.

(3) The ratio of $m_{\rm lens}/m_{\rm xray}$ increases with offset
$d$.

(4) $m_{\rm lens}/m_{\rm xray}$ depends very strongly on $R_x$. Here
$R_x$ acts like the arc radius $r_{\rm arc}$ in strong lensing,
i.e., we only measure the enclosed mass within a small region of
$R\le R_x$. When $R_x$ is very small, the offset effect is most
prominent and gives large $m_{\rm lens}/m_{\rm xray}$. Increasing
$R_x$ will reduce $m_{\rm lens}/m_{\rm xray}$. When $R_x$ is very
large (compared with $d$), the offset effect will be ``smeared
out'', and the $m_{\rm lens}$-$m_{\rm xray}$ discrepancy introduced
by the offset will vanish.

(5) The mass ratio is very sensitive to the NFW concentration, and
it increases dramatically with $c$.

(6) The mass ratio increases with the core radius, and decreases with
$\beta$ index and gas fraction. However, the mass ratio is not very
sensitive to the gas model.

\begin{figure}
\begin{center}
\protect\label{fig:theory}
\includegraphics[width=8.9cm]{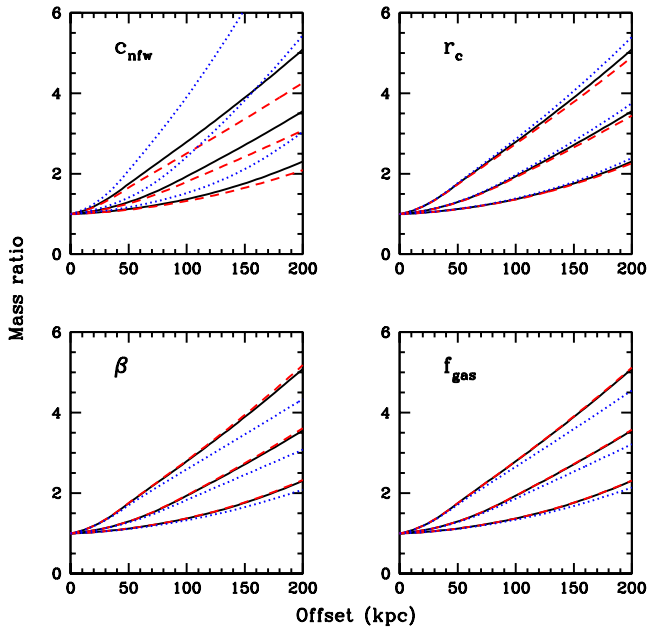}
\caption{Ratio of projected gravitational masses, as a function of
the 2-D offset $d$ and the measuring radius $R_x$. The solid curves are
the mass ratio for the fiducial model with $c=4.32$, $r_s=516 \rm kpc$, $\beta=0.65$,
$r_c=200 \rm kpc$, and $f_{\rm gas}=0.12$. The dashed and
dotted curves are for the NFW concentration $c=4.04$ and $5.13$ (top left),
the cluster core radius $r_c = 150 \, {\rm kpc}$ and $400 \, {\rm kpc}$
(top right), the $\beta$ index $\beta=0.6$ and $0.9$ (bottom left), the gas
fraction $f_{\rm gas}=0.1$ and $0.2$, respectively. The three dotted (dashed,
solid as well) curves from top to bottom in one panel correspond to $R_x=50, 100, 200
\,\rm kpc$, respectively.}
\end{center}
\end{figure}

\section{Comparison with Observational Data}
\label{sect:data}

To compare with our theoretical predictions, we compile a sample of
$27$ clusters with $48$ arc-like images, which have both strong
lensing and X-ray measurements. The clusters and their lensing and
X-ray data are listed in Table~1. For the $22$ arcs that have no
redshift information, we estimate their lensing masses $m_{\rm
lens}$ by assuming the mean redshifts of $\left< z_d \right> =0.8$
and $2.0$, respectively. The X-ray data are taken from Tucker et al.
(1998), Wu (2000), Bonamente et al. (2006), and references therein. 
The offsets between lensing and X-ray centers are taken from Shan et al. (2010).
The clusters in our table are classified as relaxed (with cooling
flow) and unrelaxed (which are dynamically unmature), from their X-ray
morphologies. The definition has been used in the literature by
Allen (1998), Wu (2000), Baldi et al. (2007), and Dunn \& Fabian (2008).

{\it Mass from strong lensing.} Assuming a spherical matter
distribution, one can calculate the gravitational mass of a galaxy
cluster projected within a radius of $r_{\rm arc}$ on the cluster
plane as
\begin{equation}
m_{\rm lens}(< r_{\rm arc})= \pi r_{\rm arc}^2 \Sigma_{\rm crit},
\end{equation}
where $\Sigma_{\rm crit}=\frac{c^2}{4 \pi G} \frac{D_s}{D_l D_{ls}}$
is the critical surface mass density, $D_l$, $D_s$ and $D_{ls}$ are
the angular diameter distances to the cluster, to the background
galaxy, and from the cluster to the galaxy, respectively. The above
equation is actually the lensing equation for a cluster lens of
spherical mass distribution with a negligible small alignment
parameter for the distant galaxy within $r_{\rm arc}$. The values of
$m_{\rm lens}$ within the arc radius $r_{\rm arc}$ are listed in
Table~1.

Allen (1998) pointed out that the use of more realistic, elliptical
mass models can reduce the masses within the arc radii by up to
$40\%$, though a value of $20\%$ is more typical. However, such
corrections are still not very significant compared with the large
discrepancies between the lensing and X-ray determined masses. We
will discuss it in more detail in the next section.

{\it Mass from X-rays.} Assuming that the intra-cluster gas is
isothermal and in hydrostatic equilibrium, the cluster mass $m(r)$
enclosed within a radius $r$ can be easily calculated from
\begin{equation}
-\frac{G m(r)}{r^2}=\frac{kT}{\mu m_p} \frac{d{\, \rm ln} n_{\rm
gas}(r)}{dr},
\end{equation}
where $T$ is the gas temperature, $n_{\rm gas}$ the gas number
density, $m_p$ the proton mass, and $\mu=0.585$ the mean molecular
weight. Here we assume that the gas follows the conventional $\beta$
model, i.e., $n_{\rm gas}(r)=n_{\rm gas}(0)(1+r^2/r_c^2)^{-\frac{3
\beta}{2}}$. In order to compare the mass measured by X-rays with
the lensing result, we need to convert this $m(r)$ (i.e., 3-D) into
the projected mass $m_{\rm xray}$ (see e.g. Wu 1994):
\begin{equation}
m_{\rm xray}=1.13 \times 10^{13} \beta \bar f\left({R \over
r_c}\right) \left (\frac{r_c}{0.1 {\rm Mpc}}\right ) \left
(\frac{kT}{1{\rm keV}}\right ) M_{\odot},
\end{equation}
where
\begin{equation}
\bar f(y) = \frac{\pi y^2}{2 (1+y^2)^{1/2} },
\end{equation}
the mass ratio $m_{\rm lens}/m_{\rm xray}$ are listed in Table~1.

Figure~3 shows the relation between the mass ratios $m_{\rm
lens}/m_{\rm xray}$ and the (scaled) offsets for our sample of $27$
clusters ($48$ arc images). It should be pointed out that the $27$
clusters in our sample have quite different sizes and masses. This
can be seen from the wide range of the cluster temperatures --- from
$4~\rm keV$ to $14~\rm keV$. Therefore, it is useful to compare the
offsets of the clusters on the same scale. We realize that the M-T
relation of clusters scales as${\rm M}\sim {\rm T}^{3/2}$ (e.g.,
Nevalainen et al. 2000; Xu et al. 2001), and that ${\rm M} \sim {\rm
R}^3$, where R is the size of the cluster. Therefore, in Figure~3,
instead of using the physical offset $d$, we use a scaled offset
which is characterized by ${\rm d_{\rm kpc}}/{\rm T_{\rm
keV}}^{1/2}$.

From Figure~3, the mass ratios $m_{\rm lens}/m_{\rm xray}$ exhibits
large dispersions --- roughly ranging from $2$ to $4$. Many clusters
have large error bars. It appears that relax clusters (marked by
crosses) have smaller $m_{\rm lens}/m_{\rm xray}$ ratios. The fact
that $m_{\rm lens}>m_{\rm xray}$ is consistent with our theoretical
predictions, and the ratio of $m_{\rm lens}/m_{\rm xray} \sim 2-4$
is also roughly consistent with our predictions as plotted in
Figure~2.

However, no strong correlation has been found between the offset and
mass discrepancies. We notice that many clusters in the sample have
very small offset values --- smaller than the errors in lensing and
X-ray measurements which are typically a few arcseconds. So these
offset values are not robustly measured themselves, and we thus
remove these data points and only focus on clusters with large
offsets of $d>10''$, as has been suggested in Shan et al. (2010).
This leaves a sub-sample of only $24$ arc images. The dashed line in
Figure~3 shows a $\chi^2$ fit to this sub-sample, which satisfies
$m_{\rm lens}/m_{\rm xray} \sim 3.24(\frac{d/100\,{\rm kpc}}{\sqrt{kT/{\rm keV}}})^{0.20}$ with
a reasonable $\chi^2=0.75$. We can find $m_{lens}/m_{xray}$ increasing slightly
with $d$.

\begin{figure}
\begin{center}
\protect\label{fig:offset-ratio}
\includegraphics[width=8.cm]{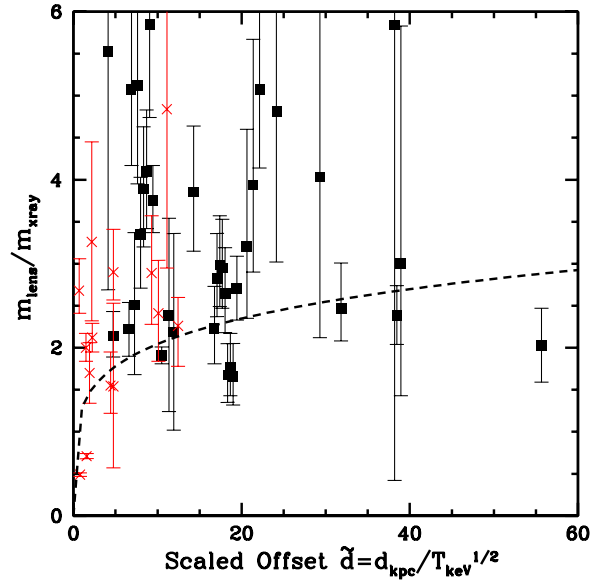}
\caption{The ratio of lensing and X-ray determined masses for our
sample of $27$ clusters ($48$ arc images). The x-label show the
scaled offset between DM and baryons. The squares denote unrelaxed
clusters, and crosses the relaxed clusters. The dashed line shows a
$\chi^2$ fit satisfying  $m_{\rm lens}/m_{\rm xray} \sim
3.24(\frac{d/100\,{\rm kpc}}{\sqrt{kT/{\rm keV}}})^{0.20}$ with
$\chi^2=0.75$ for the clusters with offset larger than $10''$.}
\end{center}
\end{figure}

\section{Discussion and Conclusions}
\label{sect:discussion}

As has been reported by Shan et al. (2010), it might be fairly
common in galaxy clusters that the X-ray center has an obvious
offset from the gravitational center. We have explored the dynamical
consequences of this lensing-X-ray offset and tried to attribute
such an effect to the  long-standing ``Mass Discrepancy Problem'' in
galaxy clusters. Our theoretical model predicts that such an offset
effect will always result in a larger $m_{\rm lens}$ than $m_{\rm
xray}$, with a typical mass ratio $m_{\rm lens}/m_{\rm xray}\sim 2$,
which is consistent with observations.

To test our model, we have compiled a sample of $27$ clusters, and
studied in detail their lensing and X-ray properties and obtained
their lensing and X-ray masses, $m_{\rm lens}$ and $m_{\rm xray}$.
The lack of strong correlation between $m_{\rm lens}/m_{\rm xray}$
and the offset $d$ suggests that the problem is more complicated. As
we have found in Section~2, $m_{\rm lens}/m_{\rm xray}$ is not only
a function of $d$, but also depends very strongly on $R_x$ (or the
arc radius $r_{\rm arc}$). Apparently, each cluster in our sample
has quite different $r_{\rm arc}$.

Probably, other mechanisms than the offset effect should play
important roles, and the lensing-X-ray mass discrepancy may not be
just from one mechanism, but a combination of many effects:

(1) The central regions of clusters may be still undergoing
dynamical relaxation, and the X-ray gas may not be in good
hydrostatic equilibrium. Therefore, large errors could be induced in
the X-ray measurement of cluster cores, especially for unrelaxed
clusters.

(2) The spherical models are too simple to reflect the real mass
distribution of clusters. The use of more realistic mass model could
reduce the lens mass within the arc radius by up to $40\%$, though
values of $\sim 20 \%$ are more typical (Bartelmann 1995; Allen
1998).

(3) The presence of substructures may complicate our simple
spherical lens model, and hence could be a main source of
uncertainties in $m_{\rm lens}$. The absence of the secondary
arc-like images in most arc-cluster systems may indicate the
limitations of the spherical mass distribution in the central
regions of clusters.

It should be noted that the mass ratios we obtained here are
slightly higher than Allen (1998) and Wu (2000) because they
unfortunately used a Hubble constant of $\rm H_0=50 \, km \,
s^{-1}Mpc^{-1}$. The use of $\rm H_0=70 \, km \, s^{-1}Mpc^{-1}$
here will of course make the mass discrepancy problem more
pronounced.

It should be noted that the gas represents only a $10\%$ perturbation
due to the small ratio of gas-to-DM in the central region, likewise the
offset of the gas is only a small perturbation (less than $10\%$) to the
otherwise concentric matter density or potential. It is unlikely to create
a factor of two difference in the lensing-derived enclosed masses within an arc.

To illustrate the lensing effect of the offset perturbation and triaxiality, we
show the critical curves in Figure~4. The solid curves indicate the
critical curve of circular NFW plus $\beta$ model without offset, the dotted curves
indicate the critical curve of elliptical NFW plus $\beta$ model with offset $d=10''$.
The square and cross denote the center of dark matter and the hot gas,
respectively. For the NFW profile, $c=4.3, r_s=516 \rm kpc$; for the $\beta$ model,
$\beta=0.65, r_c=150 \rm kpc$. We also introduce the triaxiality with the ellipticity
$e=0.15$ and position angle $\theta=30^{\circ}$. We also assume the lens and source redshifts
$z_l=0.3$, $z_s=1$. We can see that the predicted critical curves (dotted lines) have
very similar sizes as the predicted critical curves for a benchmark model (solid lines) with
the same mass DM and gas mass but in concentric spheres.

\begin{figure}
\begin{center}
\protect\label{fig:offset & triallity}
\includegraphics[width=8.0cm]{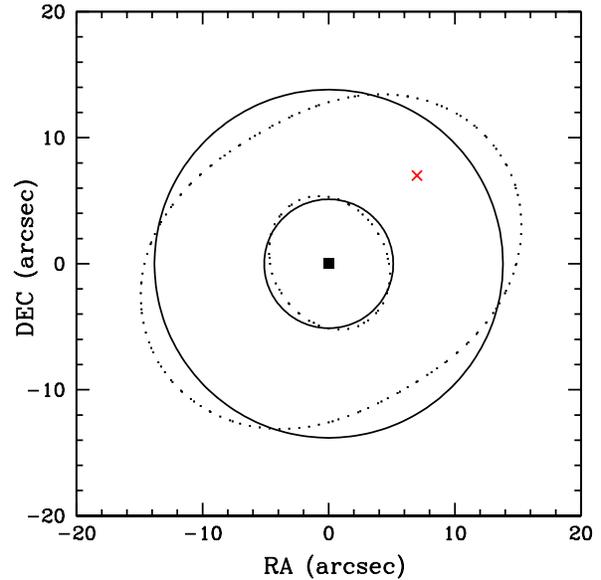}
\caption{The effects of offset and triaxiality on the critical curves. The solid curves
indicate the critical curve of circular NFW \& $\beta$ model without offset. The dotted curves
indicate the critical curve of elliptical NFW \& $\beta$ model with offset $d=10''$.
The square and cross denote the center of dark matter and the X-ray gas,
respectively. For the NFW profile, $c=4.3, r_s=516 \rm kpc$; for the $\beta$ model,
$\beta=0.65, r_c=150 \rm kpc$. The ellipticity and position angle are $e=0.15$ and
$\theta=30^{\circ}$. The lens and source redshifts are $z_l=0.3$, $z_s=1$.}
\end{center}
\end{figure}

Early studies have suggested that statistically unrelaxed clusters
have larger mass discrepancies than relaxed clusters (Allen 1998; Wu
2000; Richard et al. 2010). As Shan et al. (2010) have reported, the
clusters with large offset of $d>10''$ are all unrelaxed clusters.
If such offsets exist and are big, then they must come into play in
our dynamical studies of galaxy cluster, and should not be ignored,
especially for unrelaxed clusters.

\chapter{\flushright{\bf{Acknowledgments}}}
\flushleft{We thank Bernard Fort, Charling Tao, and Xiang-Ping Wu
for discussions, and an anonymous referee for helpful
suggestions. HYS and BQ are grateful to the CPPM for
hospitality. This work was supported by the National Basic Research
Program of China (973 Program) under grant No. 2009CB24901,
and CAS grants KJCX3-SYW-N2 and KJCX2-YW-N32.}

\onecolumn

\begin{table}
\addtolength{\tabcolsep}{-1.0pt} \scriptsize
\renewcommand\arraystretch{0.03}
\centering \caption{\rm The X-ray and lensing mass discrepancies of
$27$ clusters. For the $22$ arcs that have no
redshift information, we estimate the mean redshifts of $\left< z_d \right> =0.8$
and $2.0$, respectively. Refs A and B give the references of the lensing and X-ray data,
respectively. The last column shows the classification of the clusters:
``R/U'' means relaxed/unrelaxed.} \scriptsize
\begin{tabular}{lllllllllllllll}
\hline \hline
Cluster & $z_{\rm cluster}$ & \multicolumn{2}{c}{Offset} & $z_{\rm arc}$ & $r_{\rm arc}$ &$m_{\rm lens}$ & Ref. A$^{d}$ &kT & $\beta$ & $r_c$ & $m_{\rm xray}$ &$m_{\rm lens}/m_{\rm xray}$ & Class$^c$ & Ref.B$^{d}$ \\
\hline
        &                   &    (arcsec) & (kpc) &            &  (Mpc)          &$\times 10^{14}M_{\odot}$ &    & (keV) &       &  (Mpc)  & $\times 10^{14}M_{\odot}$ & & & \\
\hline
1E0657-56  &  0.296 & 47.4 & 209.2 &3.24 & 0.25 & 4.37  & 3,4 & $14.1^{+0.2}_{-0.2}$ & $0.62^{+0.07}_{-0.07} $  & $0.36^{+0.05}_{-0.05}$ & $ 2.15^{+0.46}_{-0.46} $ & $2.03^{+0.44}_{-0.44}$ & U & 12\\

A68$^a$  & 0.255 & 14.3 & 56.7 & 1.60 & 0.04 & 0.13 & 11 & $10.0^{+1.1}_{-0.9}$   & $0.72^{+0.04}_{-0.03}$  & $0.25^{+0.02}_{-0.02}$  & $0.08^{+0.019}_{-0.016}$  & $1.66^{+0.39}_{-0.34}$  &  U & 2\\
         &       & & &1.60 & 0.10 & 0.80  &  &                        &                            &                               & $0.47^{+0.11}_{-0.091}$    & $1.77^{+0.40}_{-0.34}$  &  & \\
         &       & & &2.63 & 0.11 & 0.94 &  &                        &                            &                               & $0.56^{+0.13}_{-0.11}$    & $1.67^{+0.38}_{-0.32}$  &      \\
         &       & & &...   & 0.211 & 4.49(3.54)$^b$ &  &                        &                            &                               & $1.69^{+0.35}_{-0.29}$       & $2.64^{+0.55}_{-0.46} (2.08^{+0.43}_{-0.36})$ &   \\
         &       & & &0.86 & 0.28 & 7.49 &  &                        &                            &                               & $2.61^{+0.51}_{-0.43}$       & $2.95^{+0.58}_{-0.48}$  &      \\
         &       & & &... & 0.27 & 7.26(5.72)$^b$ &  &                        &                            &                               & $2.47^{+0.49}_{-0.41}$       & $2.98^{+0.59}_{-0.49} (2.35^{+0.47}_{-0.39})$ &  \\
         &       & & &1.27 & 0.32 & 8.66 &  &                        &                            &                               & $3.14^{+0.60}_{-0.50}$       & $2.82^{+0.54}_{-0.45}$  &      \\
         &       & & &... & 0.12 & 1.54(1.22)$^b$ &  &                        &                            &                               & $0.65^{+0.15}_{-0.12}$    & $2.23^{+0.50}_{-0.42} (1.76^{+0.39}_{-0.33})$ &     \\

A267     & 0.230 & 9.62 & 35.3 &... & 0.12 & 1.48(1.20)$^b$ & 11 & $6.0^{+0.6}_{-0.5}$    & $0.71^{+0.03}_{-0.03}$  &  $0.19^{+0.01}_{-0.01}$  & $0.39^{+0.08}_{-0.07}$  & $3.86^{+0.78}_{-0.71} (3.13^{+0.63}_{-0.58})$  &  U  & 2\\

A370$^a$ & 0.375 & 19.9 & 102.7 &1.30 & 0.41 & 13.1 &  7 & $7.13^{+1.05}_{-1.05}$ & $0.95^{+0.75}_{-0.35}$ & $0.56^{+0.44}_{-0.26}$ & $4.34^{+4.10}_{-2.27}$ & $3.00^{+2.83}_{-1.57}$ & U & 13\\
           &     & &  & 0.72 & 0.19 & 4.09 &  &                        &                        &                        & $0.71^{+1.15}_{-0.65}$ & $5.84^{+9.54}_{-5.42}$ &    \\

A697     & 0.282 & 3.07 & 13.1 &... & 0.12 &  1.51(1.15)$^b$& 10 & $9.9^{+0.6}_{-0.6}$ & $0.61^{+0.01}_{-0.01}$ & $0.24^{+0.01}_{-0.01}$ & $0.26^{+0.21}_{-0.13}$ & $5.53^{+4.52}_{-2.84} (4.21^{+3.44}_{-2.16})$ &  U & 2\\

A773$^a$ & 0.217 & 6.43 & 22.6 &0.65 & 0.11 & 1.39 & 11 & $7.6^{+0.5}_{-0.4}$ & $0.61^{+0.01}_{-0.01}$ & $0.19^{+0.06}_{-0.06}$ & $0.37^{+0.042}_{-0.037}$ & $3.75^{+0.42}_{-0.38}$  &  U & 2\\
         &       &  & &0.40 & 0.21 & 7.08 &  &                     &                           &                               & $1.26^{+0.26}_{-0.24}$   & $5.85^{+1.19}_{-1.11}$  &    \\
         &       & & &... & 0.25 & 6.50(5.36)$^b$ &  &                     &                           &                               & $1.61^{+0.29}_{-0.26}$   & $4.10^{+0.73}_{-0.68} (3.38^{+0.60}_{-0.56})$  &  \\
         &       & & &... & 0.23 & 5.41(4.45)$^b$ &  &                     &                           &                               & $1.43^{+0.27}_{-0.25}$   & $3.89^{+0.74}_{-0.69} (3.21^{+0.61}_{-0.57})$  &  \\
         &       & & &1.11 & 0.213 & 4.34 &  &                     &                           &                               & $1.26^{+0.26}_{-0.24}$   & $3.35^{+0.68}_{-0.64}$    &   \\
         &       & & &0.40 & 0.16 & 4.14 &  &                     &                           &                               & $0.84^{+0.20}_{-0.19}$& $5.12^{+1.23}_{-1.17}$    &    \\
         &       & & &... & 0.04 & 0.18(0.15)$^b$ &  &                     &                           &                               & $0.07^{+0.023}_{-0.022}$  & $2.51^{+0.86}_{-0.83} (2.07^{+0.71}_{-0.69})$ & \\
         &       & & &0.49 & 0.23 & 7.42 &  &                     &                           &                               & $1.43^{+0.27}_{-0.25}$ & $5.07^{+0.96}_{-0.90}$  &  \\

A963$^a$ & 0.206 & 7.10 &24.0&... & 0.057 & 0.35(0.29)$^b$& 11 & $6.13^{+0.45}_{-0.30}$ & $0.51^{+0.04}_{-0.04}$ & $0.11^{+0.02}_{-0.02}$ & $0.14^{+0.038}_{-0.034}$ & $2.41^{+0.63}_{-0.57} (2.01^{+0.53}_{-0.48})$ & R & 13\\
           &       & & & 0.71 & 0.09 & 0.87 &  &                        &                        &                           & $0.31^{+0.073}_{-0.066}$       & $2.89^{+0.68}_{-0.61}$ &   \\

A1689    & 0.183 & 0.60 & 1.85&... & 0.20 & 4.5(3.8)$^b$& 8 & $9.02^{+0.40}_{-0.30}$ & $0.65^{+0.04}_{-0.02}$ & $0.14^{+0.02}_{-0.02}$ & $1.68^{+0.24}_{-0.17}$ & $2.68^{+0.38}_{-0.27} (2.29^{+0.32}_{-0.23})$ & R & 13\\

A1835    & 0.252 & 1.61 & 6.33 &... & 0.17 & 2.82(2.23)$^b$ & 11 & $9.8^{+1.4}_{-1.4}$    & $0.65^{+0.04}_{-0.04}$ & $0.08^{+0.01}_{-0.01}$ & $1.72^{+0.36}_{-0.36}$ & $1.70^{+0.36}_{-0.36} (1.35^{+0.28}_{-0.28})$ & R & 13\\

A1914    & 0.171 & 11.3 & 32.9 &... & 0.10 & 1.16(1.01)$^b$ & 10 & $9.9^{+0.3}_{-0.3}$                  & $0.90^{+0.01}_{-0.01}$ & $0.200^{+0.003}_{-0.003}$ & $0.70^{+0.036}_{-0.036}$ & $1.67^{+0.09}_{-0.09} (1.44^{+0.07}_{-0.07})$  & U & 2\\

A2204$^a$ & 0.151 & 1.20 & 3.15 &... & 0.025 & 0.08(0.07)$^b$ & 10 & $6.5^{+0.2}_{-0.2}$ & $0.48^{+0.002}_{-0.002}$ & $0.02^{+0.0003}_{-0.0003}$ & $0.11^{+0.0040}_{-0.0040}$ & $0.71^{+0.03}_{-0.03} (0.63^{+0.02}_{-0.02})$ & R & 2\\
          &       & & &... & 0.01 & 0.013(0.012)$^b$ &  &                    &                           &                                  & $0.03^{+0.0009}_{-0.0009}$ & $0.49^{0.02}_{-0.02} (0.43^{+0.02}_{-0.02})$ &  \\

A2163    & 0.203 & 44.0 & 146.9 &0.73 & 0.07 & 0.58 & 1 & $14.6^{+0.85}_{-0.85}$ & $0.62^{+0.02}_{-0.02}$ & $0.33^{+0.02}_{-0.02}$ & $0.23^{+0.033}_{-0.033}$ & $2.39^{+0.35}_{-0.35}$  & U & 13\\

A2218$^a$& 0.176 & 19.1 & 56.9 &1.03 & 0.28 & 8.60 & 11 & $7.1^{+0.2}_{-0.2}$ & $0.65^{+0.08}_{-0.05}$ & $0.25^{+0.09}_{-0.05}$ & $1.67^{+0.49}_{-0.31}$ & $5.07^{+1.49}_{-0.93}$ & U & 13\\
           &       & & & 0.70 & 0.09 & 0.89 &  &                     &                        &                           & $0.25^{+0.11}_{-0.065}$ & $3.94^{+1.73}_{-1.04}$ &    \\
           &       & & & 2.52 & 0.09 & 0.82 &  &                    &                        &                           & $0.25^{+0.11}_{-0.065}$ & $3.20^{+1.40}_{-0.85}$ &    \\

A2219$^a$& 0.228 & 11.3 & 41.2&... & 0.09 & 0.79(0.64)$^b$ & 11 & $12.4^{+0.5}_{-0.5}$ & $0.40^{+0.07}_{-0.07}$ & $0.16^{+0.08}_{-0.08}$ & $0.38^{+0.21}_{-0.21}$ & $2.19^{+1.17}_{-1.17} (1.78^{+0.95}_{-0.95})$ & U & 13\\
           &       & & & ... & 0.12 & 1.54(1.26)$^b$  &  &                     &                        &                           & $0.63^{+0.30}_{-0.30}$ & $2.39^{+1.15}_{-1.15} (1.94^{+0.94}_{-0.94})$ &    \\

A2259    & 0.164 & 16.3 & 45.9 &1.48 & 0.04 &  0.13& 10 & $5.6^{+0.3}_{-0.3}$ & $0.58^{+0.02}_{-0.02}$ & $0.14^{+0.01}_{-0.01}$ & $0.06^{+0.0089}_{-0.0089}$ & $2.71^{+0.38}_{-0.38}$ & U & 2\\

A2261$^a$ & 0.224 & 1.31 & 4.72 &... & 0.12 &   1.49(1.22)$^b$& 10 & $7.2^{+0.4}_{-0.4}$ & $0.56^{+0.01}_{-0.01}$ & $0.08^{+0.004}_{-0.003}$ & $0.71^{+0.058}_{-0.056}$ & $2.12^{+0.17}_{-0.17} (1.73^{+0.14}_{-0.14})$ & R & 2\\
          &       & & &... & 0.11 &   1.26(1.03)$^b$&  &                    &                           &                                & $0.63^{+0.053}_{-0.051}$ & $2.00^{+0.17}_{-0.16} (1.63^{+0.14}_{-0.13})$ &   \\

A2390    & 0.228 & 6.00 & 21.9 &0.91 & 0.20 & 3.8 & 10 & $11.1^{+1.0}_{-1.0}$ & $0.59^{+0.02}_{-0.02}$ & $0.16^{+0.01}_{-0.01}$ & $1.79^{+0.26}_{-0.26}$ & $2.22^{+0.32}_{-0.32}$ & U & 13\\

CL0024   & 0.395 & 13.2 & 70.4 &1.68 & 0.26 & 4.7 & 6 & $5.7^{+4.9}_{-2.1}$ & $0.48^{+0.08}_{-0.05}$ & $0.08^{+0.05}_{-0.03}$ & $1.19^{+1.22}_{-0.56}$ & $4.03^{+4.14}_{-1.91}$ & U & 13\\

MS0440   & 0.190 & 1.50 & 4.89 &0.53 & 0.10 & 1.23 & 5,10 & $5.30^{+1.27}_{-0.85}$ & $0.45^{+0.03}_{-0.03}$ & $0.03^{+0.01}_{-0.01}$ & $0.40^{+0.15}_{-0.12}$ & $3.26^{+1.19}_{-0.94}$ & R & 13\\

MS0451   & 0.550 & 12.1 & 76.8 &... & 0.23 & 7.6(3.5)$^b$ & 10 & $10.17^{+1.55}_{-1.26}$ & $0.68^{+0.13}_{-0.09}$ & $0.31^{+0.09}_{-0.06}$ & $1.64^{+0.85}_{-0.61}$ & $4.81^{+2.49}_{-1.79} (2.11^{+1.09}_{-0.78})$ & U & 13\\

MS1008   & 0.360 & 5.43 & 27.3 &... & 0.30 & 9.2(6.2)$^b$ & 1 & $7.29^{+2.45}_{-1.52}$  & $0.63^{+0.11}_{-0.07}$ & $0.23^{+0.07}_{-0.05}$ & $1.89^{+1.15}_{-0.73}$ & $4.84^{+2.95}_{-1.89} (3.29^{+2.00}_{-1.28})$ & R & 13\\

MS1358   & 0.329 & 2.79 & 13.2 &4.92 & 0.14 & 1.24 & 10 & $7.5^{+4.3}_{-4.3}$ & $0.47^{+0.02}_{-0.02}$ & $0.05^{+0.02}_{-0.01}$ & $0.82^{+0.53}_{-0.51}$ & $1.54^{+0.99}_{-0.97}$ & R & 13\\

MS1455   & 0.258 & 2.77 & 11.1&... & 0.11 & 1.22(0.96)$^b$ & 9,10 & $5.45^{+0.29}_{-0.28}$ & $0.64^{+0.04}_{-0.03}$ & $0.07^{+0.01}_{0.01}$ & $0.57^{+0.077}_{-0.067}$ & $2.14^{+0.29}_{-0.25} (1.68^{+0.23}_{-0.20})$ & U & 13\\

MS2053   & 0.580  & 10.5 & 69.1&3.15 & 0.16 &  1.41& 10 & $4.7^{+0.5}_{-0.4}$ & $0.64^{+0.04}_{-0.03}$ & $0.16^{+0.02}_{-0.01}$ & $0.60^{+0.13}_{-0.094}$ & $2.47^{+0.54}_{-0.39}$ & U & 2\\

MS2137   & 0.313 & 5.70 &26.1&... & 0.10 & 0.99(0.72)$^b$ & 9,10 & $4.37^{+0.38}_{-0.72}$ & $0.63^{+0.04}_{-0.03}$ & $0.05^{+0.01}_{-0.01}$ & $0.44^{+0.067}_{-0.094}$ & $2.26^{+0.34}_{-0.48} (1.65^{+0.25}_{-0.35})$ & R & 13\\

PKS0745  & 0.103 & 6.82 & 12.9&0.43 & 0.05 & 0.42 & 10 & $8.7^{+1.6}_{-1.2}$ & $0.59^{+0.01}_{-0.01}$ & $0.06^{+0.01}_{-0.01}$ & $0.29^{+0.075}_{-0.061}$ & $1.55^{+0.40}_{-0.33}$ & R & 13\\

RXJ1347  & 0.451 & 2.81 &16.2&0.81 & 0.28 & 8.9 & 1 & $11.37^{+1.10}_{-0.92}$ & $0.57^{+0.04}_{-0.014}$ & $0.07^{+0.01}_{-0.01}$ & $3.07^{+0.54}_{-0.35}$ & $2.90^{+0.51}_{-0.33}$ & R & 13\\

\hline \hline
\smallskip
\end{tabular}
\parbox {6.9in}
{$^a$ Multiple-arc system.}

\parbox {6.9in}
{$^b$ Arc-like image is assumed at $z_s=0.8$ $(z_s=2)$.}

\parbox {6.9in}
{$^c$ R: Relaxed, U: Unrelaxed.}

\parbox {6.9in}
{$^d$ References: (1) Allen 1998; (2) Bonamente et al. 2006; (3)
Bradac et al. 2006; (4) Clowe et al. 2006; (5) Gioia et al. 1998;
(6) Jee et al. 2007; (7) Kneib et al. 1993;
(8) Limousin et al. 2007; (9) Newbury \& Fahlman 1999;
(10) Sand et al. 2005; (11) Smith et al. 2005; (12) Tucker et al.
1998; (13) Wu 2000}

\end{table}
\normalsize

\end{document}